\documentstyle[prl,aps,floats,twocolumn]{revtex}
\begin{document}
\def\be{\begin{equation}}
\def\ee{\end{equation}}
\def\bea{\begin{eqnarray}}
\def\eea{\end{eqnarray}}
\def\E{{\rm e}}
\def\bearst{\begin{eqnarray*}}
\def\eearst{\end{eqnarray*}}
\def\peleven{\parbox{11cm}}
\def\peffec{\peight{\bearst\eearst}\hfill\peleven}
\def\pspace{\peight{\bearst\eearst}\hfill}
\def\ptwelve{\parbox{12cm}}
\def\peight{\parbox{8mm}}
\twocolumn[\hsize\textwidth\columnwidth\hsize\csname@twocolumnfalse\endcsname
\title
{Quenched Averaged Correlation Functions of the Random Magnets }
\author
{M. Reza Rahimi Tabar}
\address
{\it
$^a$ Dept. of Physics , Iran  University of Science and Technology,\\
Narmak, Tehran 16844, Iran.
\\$^b$ Institute for Studies in Theoretical Physics and
Mathematics
\\ Tehran P.O.Box: 19395-5531, Iran,\\
rahimi@theory.ipm.ac.ir}

\date{20/05/2000}
\maketitle

%%%%%%%%%%%%%%%%%%%%%%%%%%%%%%%%%%%%%%%%%%%%%%%%%%%%%%
%ABSTRACT
%%%%%%%%%%%%%%%%%%%%%%%%%%%%%%%%%%%%%%%%%%%%%%%%%%%%%%
\begin{abstract}

It is shown that the ratios of the quenched averaged three and four-point
correlation  functions of the
local energy density operator to the connected ones in the random-bond
Ising model
approach asymptotically to some $universal$ functions. We derive the explicit
expressions of these universal functions. Moreover it is shown that the 
individual logarithmic
operators have not any contribution to the connected correlation functions
of the disordered Ising model.
  
\end{abstract}
\hspace{.2in}
%\newpage
]
%{\bf 1- Introduction}
Random systems represent the spatial inhomogenuity where scale invariance
is only preserved on average but not for specific disorder realization.
The understanding of the role played by quenched impurities of the nature
of phase transition is one of the significant subjects in
statistical physics and has attracted a great deal of attention [1].
According to the Harris criterion [2], quenched randomness is a relevant
perturbation at the second-order critical point for systems of dimension
$d$, when its specific heat exponent $\alpha$, of the pure system is positive.
Concerning the effect of randomness on the correlation functions,
it is known that the presence of randomness induces a logarithmic factor
to the correlation functions of pure system [3]. Theoretical
treatment of the quenched disordered systems is a non-trivial task
in view of the fact that, one has to average the logarithm of the
partition function over various realization of the disorder in the
statistical ensemble and therefore find physical results. There are two
standard methods to perform this averaging, the supersymmetry (SUSY) approach,
and the well-known replica approach.
Recently using the replica approach it has been shown by
Cardy [4], that the logarithmic factor multiplying power law behavior are to
 be expected in the scaling behavior near non-mean field critical points. 
It is shown also that the results are valid for systems with 
short-range interactions and in an arbitrary number of dimensions. 
He concludes that in the
 limit of $n \rightarrow 0$ of replicas the theory posses of a set of
 fields which
 are degenerate (they have the same scaling dimensions)
  and finds a pair of fields which form
 a Jordan cell structure for dilatation operator and derives
 logarithmic operator in such disordered systems. 
Cardy proves that the
 quenched disordered theory with $Z=1$ can be described by logarithmic
 conformal field theory as well. The logarithmic conformal field theories (LCFT)
 are extensions of
 conventional conformal field theories, which have emerged in recent years
  in a number of interesting physical problems of condensed matter
 physics [5], string theory [6], and nonlinear dynamical systems [7].
The LCFT are characterized by the fact that their dilatation operator $L_0$
are not diagonalized and admit a Jordan cell structure.
The non-trivial mixing between these operators leads to logarithmic
singularities in their correlation functions. It has been shown [8]
that the correlator
of two fields in such field theories, has a logarithmic singularity.

\be
<\psi(r) \psi(r^{'})>\sim {|r-r^{'}|}^{-2 \Delta_{\psi}} \log|r-r^{'}| + \ldots
\ee

 In this direction we show that the quenched averaged connected correlation
 functions of local energy density field can be written in terms of
 ordinary scaling operators which
 can be constructed by the difference of energy operators in two
 different replicas. We write the connected 3 and 4-point
 correlation functions of
 energy density explicitly in terms of such ordinary operators.
 Furthermore we prove that the logarithmic operators have no
  contribution in the quenched averaged connected correlation functions
 of the local density operator. However, these operators play a considerable
 role on the disconnected ones and produce some logarithmic factors in
 the correlation functions.
 We calculate the various types of quenched averaged 3 and 4-point
 correlation functions
 of the local energy density and show that the ratios of these
 correlation functions
 to the connected ones have the specific universal asymptotic and write
 down these universal functions explicitly.

{\bf 2-Connected Correlation Functions of Random Magnets}
  
We consider a quenched random ferromagnet, for instance an Ising model, with
 random-bond disorder. Let us describe this disordered system in the continuum
 limit by the following Hamiltonian,

\be
H = H ^{0} + \int J(r) E(r) d^{d} r
\ee
where $H^{0}$ is the Hamiltonian of the renormalization group at fixed point
describing the pure Ising model. The field $J(r)$ is a quenched random
variable coupled to the local energy density $E(r)$. When the coupling $J(r)$
is independent of $x$ and not random, the above Hamitonian describes the
behavior of the
statistical model near it's critical point. For simplicity we assume that
the random variable $J(r)$ is a gaussian variable which is characterized by
its two moments $< J(r) > = 0$ and $< J(r) J(r') > = g \delta( r-r') $.
The standard procedure of averaging over disorder is to introduce replicas, 
i.e.,
$n$ identical copies of the same model for which
\be
Z^n = Tr \exp \{ -\sum_{a=1} ^n  H^{0} _a
- \int d^d r J(r) \sum_{a=1} ^n  E_a (r) \}
\ee
averaging over the disorder gives rise to the following effective replical Hamiltonaian.
\be
H_R = \sum_{a=1} ^{n} H^{0} _a - g \int \sum_{a \neq b} E_a(r) E_b(r) d^d r
\ee
we keep only the non-diagonal terms, since using the operator algebra
of the pure system one can absorb the diagonal terms into
$H^{0} _a$. The replicas are now coupled via the disorder.
The scaling dimension of coupling $g$ is $y_g = d - 2 \Delta_E$ and is 
relevant at the pure fixed point if $y_g > 0$. For small $y_g$ it is possible to
use standard perturbation theory and find the possible random fixed point.
 It is noted by Cardy that
the $n$ operator $E_a$ are degenerate at the pure fixed point and one can
decompose them into irreducible representation of permutation group $S_n$.
It has been shown that the combination $E_t = \sum _{a=1} ^n  E_a$
is a singlet ( symmetric under the permutation of the replica group)
and $\tilde E_a= E_a - \frac{1}{n} \sum _{b=1} ^n E_b $ transforms
according to an $(n-1)$-dimensional representation of $S_n$.
The fields $\tilde E_a$ satisfy the condition
$ \sum_{a=1} ^{n} \tilde E_a =0$.
The important observation is that the fields
$E_t$ and $\tilde E_a$ have the proper scaling dimensions close to
$n \rightarrow 0$ as $\Delta_{E_t} =
\Delta^{(0)} _{E_a} + \frac{1}{2} (1-n) y_g +
O(y_g^2)$
and $\Delta_{\tilde E} =
\Delta^{(0)} _E + \frac{1}{2}  y_g +
O(y_g^2)$ respectively.
It is clear that the singlet field $E_t$ becomes degenerate
with the $(n-1)$ operators
$\tilde E_a$. This is true to all orders [4].
However they do not form the basis of the Jordan cell for the dilatation operator.
To find the logarithmic pair according to [4] we define the correlation
function of $<E_t (0) E_t(r)> = A_1 $ and $< \tilde E_a (0) \tilde E_a (r)> = B_1$
and find the following relations for $A_1$ and $B_1$.
\bea
A_1 &=& n ( a - (n-1) b) \equiv n A(n) r^{-2 \Delta_E (n)} \cr \nonumber \\
B_1 &=& (1- \frac{1}{n})(a-b) \equiv
(1-\frac{1}{n}) B(n) r^{-2 \tilde \Delta_E (n)}
\eea
where $a= <E_i(0) E_i(r)>$ and $b = <E_i(0) E_j(r)>$ with $i \neq j$.
the above equations enable us to write the quenched
averaged connected two-point correlation
functions of energy density operator in terms of $a$ and $b$ in the
limit of $n \rightarrow 0$ as:
$\overline{< E(0) E(r) >}_c = a - b $ which is equal to $B(0) r^{- 2 \Delta_E}$ and
it has a pure scaling behavior. However, the correlation functions
$a$ and $b$ have the logarithmic singularities and behave as:
\bea
 < E_1(0) E_1(r) >& = & (A'(0) - B'(0) + B(0)
 \cr \nonumber \\
& -& B(0) \frac{y_g}{2} \ln r)
  r^{-2\Delta_E}  \cr \nonumber \\
 < E_1(0) E_2(r) >& = & (A'(0) - B'(0) - A(0) \frac{y_g}{2} \ln r)
r^{-2\Delta_E}
\eea
where $A(0) = B(0)$.
The prime sign in the eq. (6) means differentiating with respect to $n$.
 This means that in the limit $ n \rightarrow 0$
the fields $E_t$ and $E_a$ form a basis of Jordan cell, i.e. their two point
correlation functions behave as:
$<E_t(0) E_t(r)>=0$, $ <E_t(0) E_a(r) > = a_1 r ^{-2 \Delta_E}$
and $<E_a(0) E_b(r)> = (-2 a_1 \ln r + D_{a,b}) r^{-2 \Delta}$, where
$a_1$ and $D_{a,b}$ are some constants. As noted by Cardy the ratio
of quenched averaged two-point correlators of the energy density
operator to the connected one has a universal $r$-dependence as:
\be
\frac{ \overline{ < E(0) E(r) >} }{\overline {< E(0) E(r) >}_c} \sim
\frac{ \overline {< E(0) >< E(r)>}}{ \overline {< E(0) E(r) >}_c} \sim \ln r
\ee

To understand the structure of Jordan cell, we note that in 2D
one can define the operator $L_0$ as
\be
 L_0 =
\left( \begin{array}{ll}
\Delta_E  & 0 \\
1 &  \Delta_E \\
\end{array} \right )
\ee
so that,
$L_0 E_t = \Delta_E E_t $ and  $L_0 E_a = \Delta_E E_a + E_t$
in the limit of  $n \rightarrow 0$. Using this representation for
$L_0$ one can show that the field $E_t$ with its logarithmic
partner $E_a$ have the standard logarithmic correlation functions [9],
(see the correlation above the eq.(8) ).
We note that in 2D we have dealt with two-dimensional conformal
 field theory, relying heavily on the underlying Virasoro algebra.
 For an extention to $D$ dimensions one has to modify the
 representation of the Virasoro algebra to higher dimensions [10].
 We consider a doublet of fields (Jordan cell)
$
\Phi =
\left( \begin{array}{l}
E_t  \\
E_a \\
\end{array} \right )
$
 and note that under
 D-dimensional conformal transformation $\bf{ x} \rightarrow \bf{x'}$, we have,
 $\Phi(\bf{x})  \rightarrow \Phi'(\bf{x'}) = G ^T \Phi(\bf{x})$
 where $T$ is a two dimensional matrix which has Jordan form and
  $G=|| \frac{\partial x'}{\partial x}||$  is the Jacobian.
 For our particular case $T$ has the following Jordan form:
\be
T =
\left( \begin{array}{ll}
- \frac{2 \Delta_E}{D} &  0 \\
 1 & - \frac{2 \Delta_E}{D} \\
\end{array} \right )
\ee
 and one can show that
 the two fields $E_t$ and $E_a$, transform as:
\bea
 E_t(\bf{x'}) & = & G^{- \frac{2\Delta_E}{D}} E_t (\bf{x})  \cr \nonumber \\
 E_a(\bf{x'}) & = & G^{- \frac{2\Delta_E}{D}} ( \ln (G) E_t (x) + E_a (x) )
\eea
This expresses that the top-field $E_t$ always transforms as an ordinary
scaling operator. It can be verified that the correlation functions of
fields $E_t$ and $E_a$ have the standard
D- dimensional logarithmic conformal field theory structure [9,10].
 Using the above results, it is evident that the dimension
of field-difference $ E_a - E_b $ with $a \neq b$ is $\Delta_E$ and it
transforms as an ordinary operator under the scaling transformation.
The intresting observation is that 
%is that the individual logarithmic operator $E_a$
%do not contribute to the $connected$ quenched averaged
%correlation functions.
%Instead on
 the connected averaged correlation functions depends on the
difference fields $ E_a - E_b $ only and therefore they behave as the
ordinary correlation functions. 
For instance in the following we write
the
connected quenched averaged 2,3 and 4-point functions of local energy
density in terms of the
field-difference operators explicitly,
\bea
\overline {< E(1) E(2) >}_c &=& \frac{1}{2} < (E_a-E_b)_{(1)} (E_a -E_b)_{(2)}>
\eea
\bea
\overline{< E(1) E(2) E(3) >}_c& = &
< (E_a-E_b)_{(1)} \cr \nonumber \\
&& (E_a-E_c)_{(2)} (E_a-E_b)_{(3)}>
\eea
\bea
&& \overline{< E(1) E(2) E(3) E(4)>}_c = \cr \nonumber \\
&&< (E_a-E_b)_{(1)} (E_a-E_c)_{(2)}  (E_a-E_d)_{(3)} (E_a-E_b)_{(4)} > \cr \nonumber \\
&&-\frac{1}{2} < (E_a-E_b)_{(1)} (E_c-E_d)_{(2)}  (E_c-E_d)_{(3)} (E_a-E_b)_{(4)} > \cr \nonumber \\
&&-\frac{1}{4} < (E_a-E_b)_{(1)} (E_c-E_d)_{(2)}  (E_a-E_b)_{(3)} (E_c-E_d)_{(4)} > \cr \nonumber \\
&&-\frac{1}{4} < (E_a-E_b)_{(1)} (E_a-E_b)_{(2)}
 (E_c-E_d)_{(3)} (E_c-E_d)_{(4)} >  \nonumber
\eea
where the last equation has only 15 independent terms.

To confirm this prediction also one can directly show that the
quenched averaged connected correlation functions
have a pure scaling behaviour which is determined by 
ordinary scaling operators
and the logarithmic operators $E_a$ do not change
its behavior. This can be verified directly for the 
quenched averaged
connected 3-point
correlation function of energy density.

We are interested in deriving exactly the various 3-point quenched averaged
functions as $ \overline{< E(1) E(2) E(3) >}$,
$ \overline{<E(1) E(2)><E(3)>}$ and
$\overline{<E(1)><E(2)><E(3)>}$, which can be written in terms of the replica
correlation functions $ < E_1 (1) E_1 (2) E_1 (3) > = a $
$ < E_1 (1) E_1 (2) E_2 (3) > = b $ and
$ < E_1 (1) E_2 (2) E_3 (3) > = c $, respectively.
One can derive the correlation functions $a$, $b$ and $c$ by means
of 3-point functions of $E_t$ and $\tilde E_a$ as follows:
\bea
&&< E_t(1) E_t(2) E_t(3) > = na+3n(n-1)b \cr \nonumber \\
&&n(n-1)(n-2) c \equiv n A_1
\eea
\bea
&&< \tilde E_a(1) \tilde E_a(2) E_t(3) > = n_1 a +( n_1 ^2 (n-1) \cr \nonumber \\
&& -4n_1 ^2 + \frac{1}{n^2} (n-1)^2 + \frac{2}{n^2} (n-1)(n-2))b
\cr \nonumber \\
&+& ( -\frac{2}{n}
n_1 (n-1) (n-2) + \frac{1}{n^2} (n-2)^2 (n-1) ) c
\cr \nonumber \\ &&\equiv (1- \frac{1}{n}) B_1
\eea
and finally,
\bea
&&< \tilde E_a(1) \tilde E_a(2) \tilde E_a(3) > = (n_1^2 - \frac{n-1}{n^3}) a
 \cr \nonumber \\
&& (-3 n_1^2 \frac{n-1}{n} -\frac{3}{n^3} (n-1)(n-2) + \frac{3}{n^2} n_1 (n-1))b
  \cr \nonumber \\
&& + ( \frac{3}{n^2} n_1 (n-1) (n-2) - \frac{1}{n^3} (n-1)(n-2)(n-3) ) c
 \cr \nonumber \\
 &&\equiv (1- \frac{1}{n}) (1- \frac{2}{n}) C_1
\eea
where $n_1 = (1 - \frac{1}{n})$ and $A_1$, $B_1$ and $C_1$ are pure scaling
functions of variables $r_{i,j}$.
To derive the above equations we use the replica symmetry and
symmetry of the
various types of 3-point correlation functions under interchanging of
positions.
We note that replica symmetry leads to
have $ <\tilde E_a(1) E_t(2) E_t(3) > = 0 $ and therefore,
dose not give any new relationship
between $a$, $b$ and $c$. Using the above equations, it can be found that
the correlation functions $a$, $b$ and $c$ are as follows:
\bea
a &=& \frac{3nB_1 -3nC_1 + n^2 C_1 +A_1 - 3B_1 +2 C_1}{n^2} \cr \nonumber \\
b &=& \frac{nB_1 - n C_1 +A_1 - 3B_1 +2 C_1}{n^2} \cr \nonumber \\
c &=& \frac{A_1 - 3B_1 +2 C_1}{n^2}
\eea
 
Using the above equations we can show that the
connected quenched averaged 3-point function behaves as:
\bea
\overline{< E(1) E(2) E(3) >} &=& 2 c + a -3b  \cr \nonumber \\
&=& C_1
\eea
which is a scaling function and confirms the observation that the
logarithmic operators (individually)
have no role in the connected quenched averaged
correlation functions.
In addition one can derive the correlation functions
$< E_i (1) E_j (2) E_k (3)> $
for given $i$, $j$ and $k$
in
the limit of $n \rightarrow 0$ and show that they have the following form:
\bea
<E_i(1) E_j(2) E_k(3)> & = & [ \alpha_{ijk} - \beta_{ijk} D_1 \cr \nonumber \\
&+& \gamma_{ijk} (4D_2 - D_1 ^2)] f(1,2,3)
\eea
where $f(1,2,3) = ( r_{12} r_{13} r_{23} ) ^{-2 \Delta_E}$
 , $D_1 = \ln(r_{12} r_{13} r_{23})$ and
 $D_2 = \ln r_{23} \ln r_{13} + \ln r_{13} \ln r_{12} + \ln r_{23} \ln r_{12} $.
It can also be shown that the ratio of various symmetrized 3-point 
functions to the 
connected one behaves asymptotically as a $universal$ function
\be
\frac{1}{3}(4 D_2 - D_1^2).
\ee

We generalize the above calculations to derive the
various type of 4-point correlation functions and show that
the ratio of the various disconnected to the connected one
have the following universal asymptotic:

\be
\sim \frac {1}{36} [ O_1 ^3 - 6 O_2 - 3 O_3 - 12 O_4 - 18 O_5]
\ee

where $O_1= \ln ( r_{12} r_{13} r_{14} r_{23} r_{24} r_{34})$,
$O_2= (\ln r_{ij} \ln r_{kl} ^2 + \cdots)$ with $ i\neq j \neq k \neq l$,
$O_3 = (\ln r_{ij} \ln r_{ik} ^2 + \cdots)$ with $ i\neq j \neq k$,
$O_4 = (\ln r_{ij} \ln r_{kl} \ln r_{lj} + \cdots)$ with $ i\neq j \neq k \neq l$,
and finally
$O_5 = (\ln r_{ij} \ln r_{ik} \ln r_{il} + \cdots)$ with $ i\neq j \neq k \neq l$

In summary, in this paper we have studied the correlation functions of
disordered random magnets [11] and obtain the various types of
3 and 4-point quenched averaged
correlation functions. One can check directly that these different 
types of the 3 and 4-point
correlation functions have the general
property of a logarithmic conformal field theory that the logarithmic partner
can be regarded as the formal derivative of the ordinary fields (top field)
with respect to their conformal weight [9]. In this case, one can
consider the $E_a$ fields as the derivative of $E_t$ with respect to $n$ . 
We emphasise that the derivative with respect to scaling weight
can be written in terms of the derivative with respect to $n$.
These properties enable us to calculate any N-point
correlation function containing the logarithmic field  $E_a$ in terms
of the correlation functions of the top-fields. The general expression of the
correlation functions of the LCFT's are given in ref.[9] and here 
we determine the unknown constants in the logarithmic correlation
functions in terms of details of the random-bond Ising model.
It is noted that the formal derivations with respect to
scaling dimensions can not predict the unknown  constants in the
quenched averaged correlation functions of the local energy density operators.
The constant depends on the detail of the statistical model. 
%and its universality class.
We have shown that the individual logarithmic operators $E_a$
do not have any contribution to the quenched averaged
connected correlation functions of the energy density. We 
also obtain that the connected
correlation functions can be written in terms of the difference fields
which transform as an ordinary scaling operator.
However they will play a crucial role to the disconnected averaged
correlation functions.
Also we find that the ratio of the various
types of 3 and 4-point quenched averaged
correlation functions to the connected ones have a universal asymptotic
behavior and give their explicit form.
These predictions can also be investigated numerically.
Our analysis are valid in all dimensions as long as the dimension is
below the upper critical dimensions.To derive the above results we
have used the replica symmetry. 
Any attempt towards the breaking of this symmetry will
change completely the above picture and produces more than
one logarithmic fields in the block and produces higher order
logarithmic singularities [9].

These results can be easily generalized to other problem such as
polymer statistics, percolation and random phase sine-Gordon model etc.

We would like to thank John Cardy for his useful comments and 
A. Aghamohammadi and R. Asgari, B. Davoudi and J. Davoudi
for thier useful discussions. This paper is dedicated to Dr. A. M. Zaker,
associate professor of Physics Department,
who deceased last may 2000.

\vskip -.5cm
%\newpage

\end{document}